# Towards a Holistic Approach to Designing Theory-based Mobile Health Interventions

Yunlong Wang[1], Ahmed Fadhil[2], Jan-Philipp Lange[3], Harald Reiterer[1]

[1]HCI Group, University of Konstanz, Konstanz, Germany, {firstname.surname}@uni.kn

[2]Centro Ricerche GPI, Fondazione Bruno Kessler (FBK-irst), Trento, Italy, fadhil@fbk.eu

[3]Social and Health Sciences Group, University of Konstanz, Konstanz, Germany, jan-philipp.lange@uni-konstanz.de

## Abstract

Increasing evidence has shown that theory-based health behavior change interventions are more effective than non-theory-based ones. However, only a few segments of relevant studies were theory-based, especially the studies conducted by non-psychology researchers. On the other hand, many mobile health interventions, even those based on the behavioral theories, may still fail in the absence of a user-centered design process. The gap between behavioral theories and user-centered design increases the difficulty of designing and implementing mobile health interventions. To bridge this gap, we propose a holistic approach to designing theory-based mobile health interventions built on the existing theories and frameworks of three categories: (1) behavioral theories (e.g., the Social Cognitive Theory, the Theory of Planned Behavior, and the Health Action Process Approach), (2) the technological models and frameworks (e.g., the Behavior Change Techniques, the Persuasive System Design and Behavior Change Support System, , and the Just-in-Time Adaptive Interventions), and (3) the user-centered systematic approaches (e.g., the CeHRes Roadmap, the Wendel's Approach, and the IDEAS Model). This holistic approach provides researchers a lens to see the whole picture for developing mobile health interventions.

## Introduction

### Why Health Behavior

According to the County Health Rankings [1], variation in health can be accounted for by health behaviors (30%), clinical care (20%), social and economic factors (40%), and physical environment (10%). Increasing evidence shows that lifestyle-related behaviors, such as diet, exercise, sleeping, emotion, and smoking play an important role in people's health. Chronic diseases caused by unhealthy behaviors and habits are among the leading causes of death [2].

### The Potential of Mobile Health

The WHO's Global Observatory for eHealth (GOe) defined *Mobile Health* (*mHealth*) as a medical and public health practice supported by mobile devices and platforms, such as smartphones, patient monitoring devices, personal digital assistants (PDAs) and other wireless devices [3]. We follow this definition with emphasis on the smartphone as the core component, due to its availability to a larger population compared to other mobile devices and its suitability to deliver behavior change interventions. Having important features like portability, rich integrated sensors, privacy, and great computing power, smartphones hold great potential to improve the monitoring and treatment of chronic diseases by providing in-situ or context data and tailored timely interventions.

Mobile devices have been the dominant entrances to obtain the fast-grown information. According to a study conducted in the USA [4], the time adults spend with mobile digital media per day is now significantly higher (51% of total spend time) compared to the desktop (42%). Although smartphones are deeply embedded in our daily life for communication, entertainment, and socialization, they have limited applications in health and wellness domain.

## mHealth Intervention Design Challenges

As conducting research in the context of human health is complex, applying this to mHealth research makes it more difficult. Firstly, health related data, such as patient data in general, daily diet or activity data, behavioral and emotional data are still hard to obtain and not available openly [5]. Secondly, the current development of mHealth technology often disregards the interdependencies between technology, human characteristics, and the socioeconomic environment, resulting in technology that has a low impact in healthcare practices [6], which makes condition controlled experiments less suitable for mHealth research. There is a need for a holistic approach to design and develop mHealth technologies which consider the complexity of health-care and the rituals and habits of patients and other stakeholders. Lastly, health intervention design may need knowledge and experience from several disciplines, such as public health, behavioral science, human-computer interaction, and mobile computing.

A growing body of research has been devoted to health interventions and suggested that behavioral theory based interventions are more effective than those with no such foundations [7–9]. While some works have contributed to our understanding of user-centered design and how to develop behavior change interventions [10–13], others contributed more to provide solid behavioral theories and cognitive analysis of health interventions. User-centered design draws on multiple sources of knowledge to support creating systems which are based on users' capabilities, and the task being conducted. While behavior theories provide the foundations for user's cognitive load and routine. Researchers from both sides are more likely to miss information from the other domain due to lack of exposure and expertise in the domain. However, all these approaches suggest that we should consider variation and similarity in the contexts, people, and tasks that characterize different design situations and settings.

## Aims

We aim to build a holistic approach to aid the research of mHealth intervention design using existing theories and models. In this paper, we first present the model of our holistic approach and then demonstrate the relevant theories and models in three categories: behavioral theories, technological frameworks for the implementation of mHealth interventions, and system approaches to designing behavior change products in the context of mHealth.

# The Holistic Approach

The comprehensive model is shown in Figure 1. In our holistic approach, the whole process starts with the *understanding* of human mind, behavior change, and the specific problem, followed by determining the target user group, target behavior, and outcomes. Behavioral theories, as the basis of the whole process, should be applied in the first two steps and should be always kept in the process while following steps to guide the refinement process. Then the technological model is used for *designing* the intervention components. The interventions are *evaluated and refined,* accordingly. The refinement may require several iterations based on the evaluation results and design aims in practice. This holistic approach combines behavioral theories and technological models into a user-centered design process.

Compared to other organizing frameworks [5–7], with the proposed holistic approach, we put emphasis on categorizing and integrating components from relevant research. Our approach aims to bring together researchers from different disciplines (i.e., behavioral psychology, public health, human-computer interaction) by illustrating the connections of knowledge in these fields.

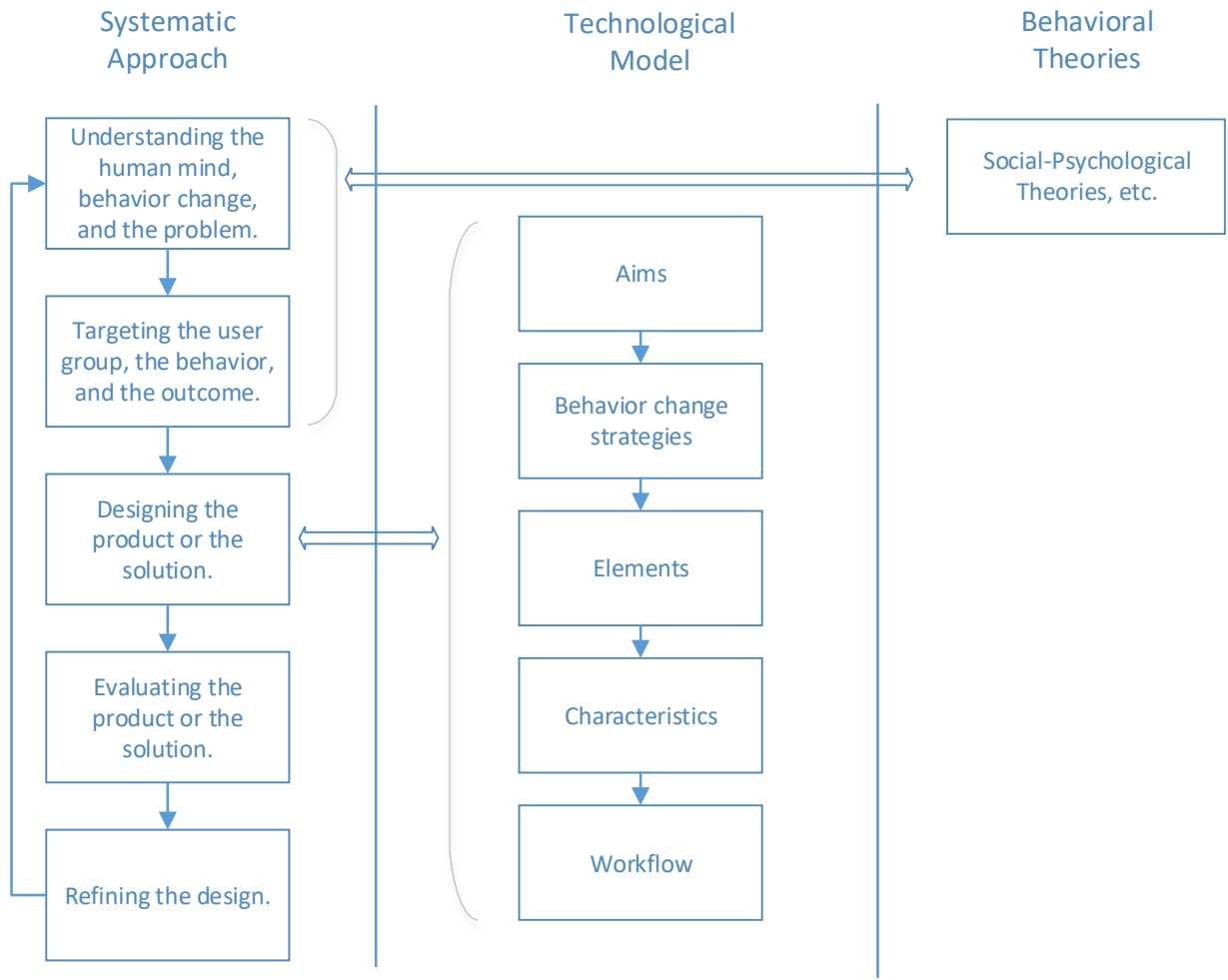

Figure 1: The Holistic Approach Model.

## Scenario

To better illustrate our approach, we apply it to a scenario where a mobile application is designed to decrease the sitting time at work for office workers in an IT company.

*The company has deployed a sit-stand desk in the office, but most employees do not use it regularly. The mobile application designer, Cindy, leads a team including a behavioral scientist and a mobile application developer. Given the target people and behavior, they first try to understand the problem by interviewing and performing a questionnaire-based survey based on Theory of Planned Behavior and Health Action Process Approach. They find that risk perception and maintenance self-efficacy are the dominant factors that cause the low adoption of the sit-stand desk. "I know it (standing) is helpful, but I just cannot persist in using the sit-stand desk, I always forget to use it", says an office worker. After understanding the problem, they decide that the proximal outcome should be increasing the frequency of using the sit-stand desk in standing mode, which is measured by a proximity sensor. Cindy then starts to design the workflow and the GUI of the application. A notification will pop up when the app detects the user entering the company and lead the user to the screen showing a summary of yesterday's sitting time summary (i.e., self-monitoring of behavior) and potential sedentary lifestyle related diseases (i.e., threat). If the app detects that the table has been in sitting mode for an hour, a notification will show up which suggests to taking a break and turn to stand up (i.e., reminder). The programmer implements the functions based on Cindy's design and tests it in a pilot study with three participants. In the evaluation, they find two participants would like to know their colleagues' sitting time. The behavioral scientist also suggests adding the social support and competition function to the app. Based on this feedback, they refine the app and deliver it to all the target workers in the company. One month later, the participants' sitting time significantly decreases.*

In this scenario, the designer goes through all the steps of the holistic approach, and finally create an effective mHealth application. Following, we will illustrate the details in this approach and show the underlying rationale.

## Behavioral Theories
### Why behavioral theories?

By behavioral theories, we refer to the social-psychological theories of behavior change (e.g., the Social Cognitive Theory and the Transtheoretical Model of Change), which have been developed and evaluated for decades, even before the age of the internet and mobile technology. As we mentioned, increasing evidence shows that theory-based behavior change interventions are more effective than others [7–9]. The behavioral theories are not only used for understanding and predicting health behavior, more importantly, they are imperative to increase the effectiveness of designing health interventions. For example, if a person does not have enough risk perception (an element in behavior change theories) of prolonged sedentary behavior, he may not change his sitting habit even if he is provided with a standing desk at work. To apply these theories, a big body of research is devoted to study behavioral intervention techniques to bring theories into applications. Interventions could range from "simple" to more complex. For example, a simple intervention is encouraging individuals quit smoking (simple in the goals at least; smoking cessation is quite complex to achieve) [8,9], whereas a more complex intervention would target individuals with several health risk factors and encourage a variety of behavioral changes, such as eliminating cigarette smoking, lowering consumption of fatty foods, and reducing overall caloric intake [8,9]. The idea of behavioral change is to change the actions of people, rather than to act on individuals passively [8,9]. Applying behavioral change techniques is the translation of a behavioral theory framework or method into a specific context, population, or culture [10]. An application of behavioral change technique is the practical incarnation of the method in an intervention. For example, one intervention can use modeling by using a vignette, whereas another intervention can use the same theoretical method (i.e., modeling), but in a completely different incarnation, for example by organizing peer education.

## Which Behavioral Theories?

Glanz et al. [11] illustrated the most frequently used behavioral theories published before 2010: the Social Cognitive Theory (SCT) [12], the Transtheoretical Model of Change (TTM) [13], the Health Belief Model (HBM) [14], and the Theory of Planned Behavior (TPB) [15]. Davis et al. [16] also identified 82 social-psychological theories of behavior change, among which the most frequently used theories are the TTM, the TPB, the SCT, the Information-Motivation-Behavioral-Skills Model, the HBM, the Self-determination Theory [17], the Health Action Process Approach [18], and the Social Learning Theory [19].

Several common theories appeared in other literature reviews [20,21]. Based on these reviews, we selected five to further review in this section, namely the SCT, the TTM, the HBM, the TPB, and the HAPA. Based on the category method in [18,22], we divide these theories into two groups, continuum models (the SCT, the HBM, and the TPB) and stage models (the TTM, and the HAPA). Besides the four most used theories, we include HAPA because it integrates the characteristics of both categories and it bridges the intention-behavior gap using mediator planning [23]. We differentiate continuum models and stage models because of their distinct assumptions on the behavioral change process. We believe with mobile technologies we can support the evaluation of these models, because more context information will be provided by this technology to detect and extract information (i.e., users' stages of behavior change to deliver more accurate interventions according to the models). Finally, we encourage the mobile health intervention designers to report their model as explicitly as possible to empower further systematic reviews.

## Review method

To provide a clear method of selecting and integrating these theories into the behavior change intervention design process, we review each theory by first describing it with a model diagram, and then show the evidence supporting or refuting it.

## Continuum Models of Behavior Change

In the continuum models of behavior change, it is assumed that a person's behavior is the outcome of conscious intention and the intention is influenced by a range of predictor variables [18], including perceived barriers, social norms, disease severity, personal vulnerability, and perceived self-efficacy, among others. In this section, the SCT, the HBM, and the TPB are reviewed.

### The Social Cognitive Theory (SCT)

Self-efficacy, as the core predictor of behavior, is people's belief in their ability to perform a target behavior [12,24]. Self-efficacy and outcome expectation affects behavior directly and encourage goals, which then affects behavior. Moreover, social-structural factors, such as financial state and environmental system also impact on individuals' goal setting. The relationship between these constructs is shown in Figure 2. Depending on the specific target behavior, adaptive questionnaires should be designed to assess the determinants based on existing guidelines [24] or examples [25].

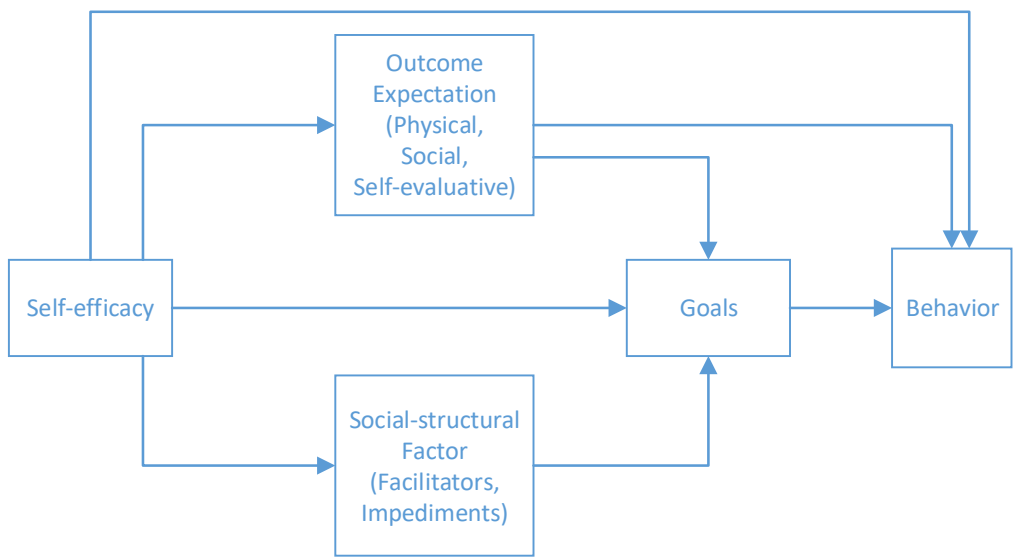

Figure 2: The SCT, adapted from [26].

A systematic review [26] shows that the SCT is an effective framework to explain Physical Activity (PA) behavior. The meta-analysis reveals that the SCT constructs accounted for 31% of the variance in PA, while only self-efficacy and goals are consistently associated with PA. In terms of the intervention's potential, another systematic review of SCT-based interventions for cancer survivors [27] implies that SCT-based interventions targeting diet or physical activity are safe and result in meaningful changes to diet and physical activity behavior that can result in health improvements [28–30]. This theory explains behaviour learning through observation and social contexts. Centred on the belief that behaviour is a context of the environment through psychological processes. Some of the behavioural change techniques that are commonly applied in this theory include, provide information on consequences, prompt barrier identification, prompt intention formation, provide general encouragement, set graded tasks, provide instruction, model or demonstrate the behaviour [31].

### The Health Belief Model (HBM)

According to the HBM, a person will take a health-related action (e.g., taking regular exercise) if that person: 1) Perceives threats, including perceived susceptibility (i.e., one's perception of the risk or the chances of contracting a health disease or condition) to illness or health problems and perceived severity (i.e., the degree one judges a disease or condition is serious); 2) Comes to a positive evaluation regarding perceived benefits (e.g., taking up regular exercise can help to reduce weight) and perceived barriers (e.g., the new exercise regime is time consuming and is difficult to keep up); 3) Perceives ability (self-efficacy), i.e. being capable of successfully performing a recommended health behavior (e.g., he/she can take up regular exercise); 4) receives some cues to action (e.g., an advertisement) [11,14,24]. As shown in Figure 3, demographic factors and psychological factors are believed to have an indirect impact on the readiness to the target behavior in the HBM. To create a questionnaire for measuring the HBM constructs, one can follow Champion's example [32].

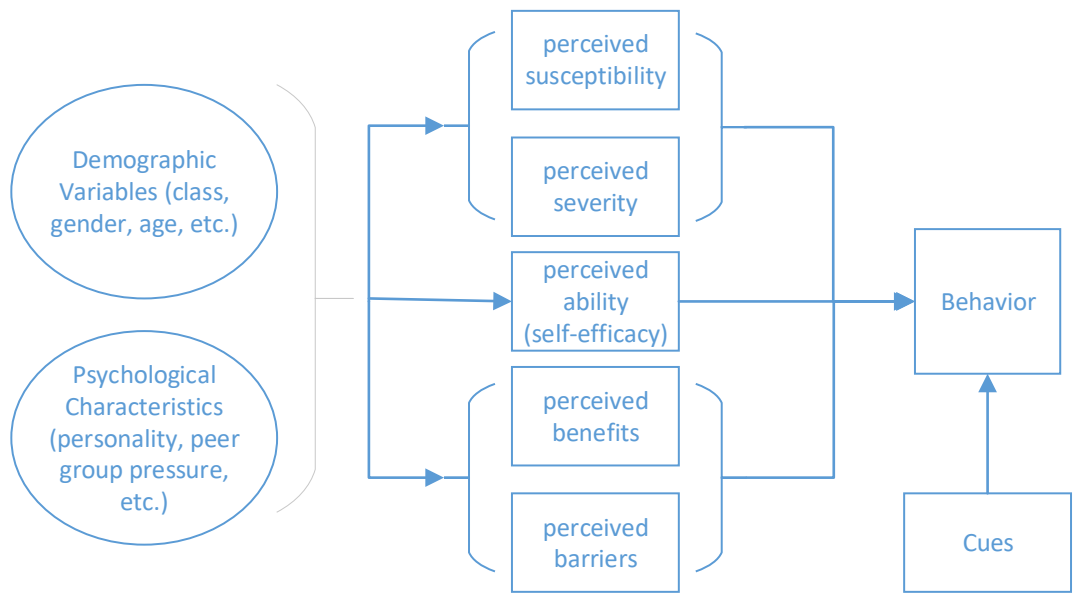

Figure 3: The HBM, adapted from [24].

The related systematic reviews [33–35] show that perceived benefits and barriers are the strongest predictors for explaining health behavior. The most recent systematic review of the HBM-related interventions in improving adherences suggests that evidence for endorsing the model for use in adherence-enhancing interventions is weak because no consistent relationship between HBM construct addressed and intervention success is found [36]. Besides, the relationships between perceived susceptibility and severity as well as perceived benefits and barriers are not well-defined in this model, which limits the applicability of the HBM and weakens the power of meta-analyses of the HBM. The model focuses on the attitudes and beliefs of individuals. The HBM is commonly used in many health actions, such as sexual health actions (condom use). The most commonly used behavior change techniques when mapping this model to an application are individual's knowledge, plan social support or social change, and educational information through behavior [9]. In addition, among the factors to consider when applying this model are perceived susceptibility/severity/benefits/barriers, readiness to act, cues to action, and self-efficacy [9].

### The Theory of Planned Behavior (TPB)

The TPB states that attitudes toward the behavior, subjective norms, and perceived behavioral control together shape an individual's behavior and behavioral intentions, as shown in Figure 4. Attitudes, subjective norms (i.e., the perceived social pressure to engage or not to engage in a behavior) and perceived behavioral control are assumed to be based on the strength and evaluation of accessible behavioral, normative and control beliefs [37]. Perceived behavioral control is hypothesized to influence behavior directly (rather than indirectly through intention) and depends on the degree of *actual control* over the behavior. To apply the TPB, one can refer to Ajzen's description of constructing the TPB questionnaire [38].

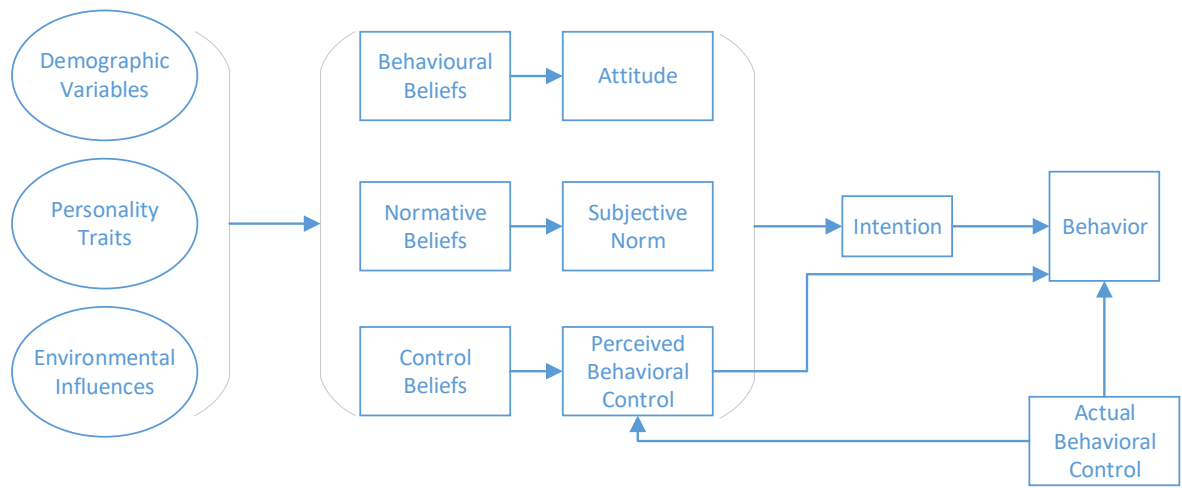

Figure 4: the TPB, adapted from [24].

Consistent evidence shows that attitude has a strong association with intention, followed by subjective norm and perceived behavior control for the prediction of dietary behavior [39,40] and alcohol consumption [41]. However, the effectiveness of TPB-based intervention has not been well evaluated [37,42]. The model aims to predict the specific plan of an individual to engage in a behavior (time and place), and apply to behaviours over which people can enact self-control over. Among the behavioural change techniques applicable to this model are providing information on consequences, providing information about others' approval, prompting intention formation [31].

## Stage Models of Behavior Change

Stage models of behavior change mostly focus on understanding readiness to make a change, appreciating barriers to change, and helping to anticipate relapses to improve patient satisfaction and lower the frustration during the change process. The models in this category follow a certain pattern by dividing the process of behavior change into discrete stages of change [43]. A central assumption of those models is that the different stages are characterized by different combinations of determinants, which are unique for the respective stage. Based on their preparedness and level of motivation to change, participants can be assigned to a stage of change. This approach allows tailoring interventions based on the hypothesized stage-specific needs of a targeted user. In the following, we discuss the relevant models and frameworks that apply a stage based approach.

### The Transtheoretical Model (TTM)

The TTM, also known as stages of change model, is focused on depicting persons as being in a process of change [13,44]. For example, people might start from the "Precontemplation" stage at which they are uninterested, unwilling to make a change ("No, and I do not intent to stop smoking within the next six months"), to the "Contemplation" stage where one starts considering a change ("No, but I intend to stop smoking within the next six months"), to deciding and preparing to make a change ("No, but I intend to start within the next 30 days"). Then, determined action is taken ("Yes, but for less than six months") and the new behavior might be maintained ("Yes, for more than six months"). This model, as shown in Figure 5, divides the behavior change process into five stages, namely *precontemplation, contemplation, preparation, action, and maintenance.* Depending on the current stage of change, a different strategy could be applied accordingly to make the intervention effective. For example, if an individual is in the preparation stage, it makes more sense to support them to develop realistic goals and timeline for behavior change and provide them positive reinforcement to move them into the action stage.

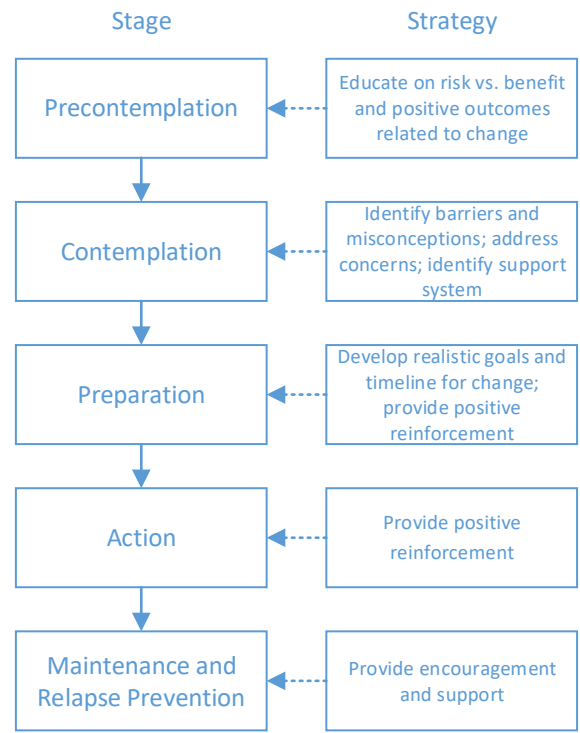

Figure 5: The TTM, adapted from [45].

Researchers argue that interventions based on the TTM have limited effect [46] because validated staging algorithms and activity change determinants are not included in the TTM. Some studies [47] have shown that many people think of themselves as complying with recommendations for complex behaviors, such as low fat intake, fruit and vegetable consumption, as well as physical activity, while their actual behavioral patterns are not in line with the recommendations. Since staging algorithms are usually based on self-assessment, these people would be regarded as being in the maintenance stage, while in fact their actions are not in line with the recommended activity levels and they show no motivation to change. On the other hand, transitions from a pre-action stage of change to an action stage of change did not necessarily coincide with actual behavior change in behavior change interventions [46,47]. This theory incorporates aspects of the integrative biopsychosocial model [48]. Among the behavioural change interventions used when applying this theory are anticipated barriers, skill development through practice, consequences and benefits, getting help, acquisition of skills, using rewards [16]. Some applications of this model into health context include, physicians counselling smokers, Type 2 Diabetes Patients counselling.

### The Health Action Process Approach (HAPA)

The HAPA constitutes an implicit stage model [18] as it suggests a distinction between the motivational processes that lead to a behavioral intention and volitional processes that lead to the actual health behavior. Therefore, interventions with different focuses can be designed respectively. On the other hand, some researchers do not regard the HAPA as a stage model like TTM because 1) If the continuum version of the HAPA is regarded as an implicit stage theory, then the TPB and several other widely used social cognition models would also have to be regarded as implicit stage theories, and we would lose the clear and useful distinction between (explicit) stage theories, such as the TTM, and continuum theories, such as the TPB, which is fundamentally different in structure. 2) The causal processes involved in motivation and volition are part of the same causal model that motivation causes volition [23] instead of a transition from motivation to volition. As illustrated in Figure 6, the HAPA bridges the intention-behavior gap by introducing planning as a mediator. In addition, phase-specific self-efficacy is emphasized to gain a better understanding of the relationship between self-efficacy and other constructs.

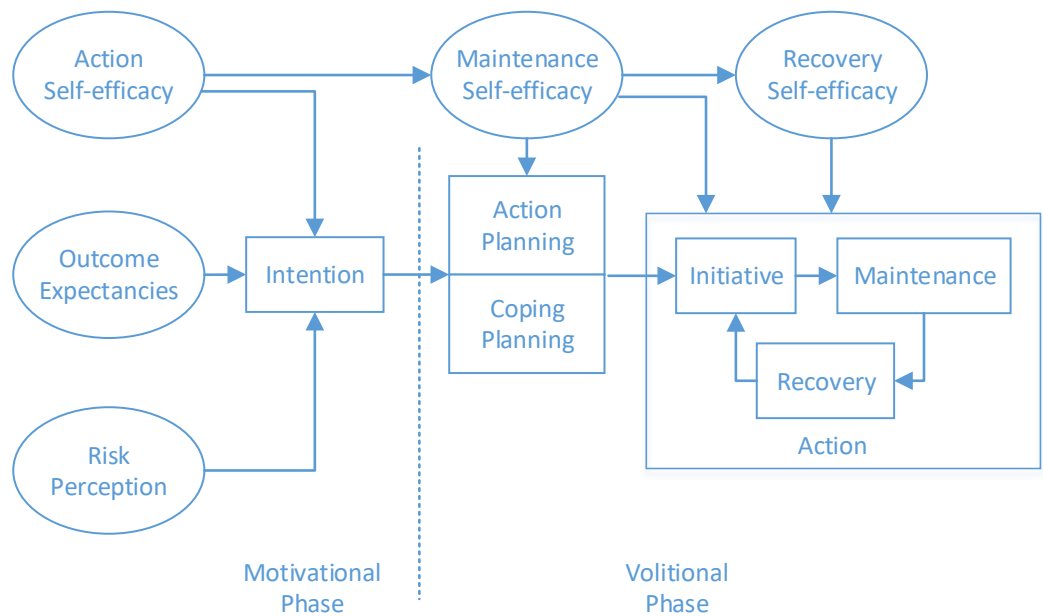

Figure 6: the model of HAPA, adapted from [18].

To our knowledge, there is only one systematic review of intervention studies using HAPA [49]. There were seven trials with appropriate data reported that focused on weight-related outcomes were included in the meta-analysis of this review. The result showed that HAPA-like (includes self-efficacy plus four other defined HAPA components, at least one from the motivational phase and one from volitional phase) interventions, resulted in statistically and clinically significant weight loss. The behavioural change techniques commonly applied in this model include, goal oriented technique for eliciting and strengthening intrinsic motivation for change, and planning ahead [50]. Some applications of the HAPA model in the context of health include physical activity training and dust mast wearing [50].

## Discussion

The variety of constructs in these theories help us to gain a broad understanding of health behavior change. As no theory mentioned can fit all behavior change situations, the research aiming to maximize effectiveness should not be limited to one specific theory, but rather utilize multiple theories to identify and focus on the most influential factors for the target person or population [51]. It is important to note that researchers from both behavioral theories and user-centered design should be familiar with each theory's unique contribution to the intervention or behavior change to avoid redundancy and arbitrariness when combining theories.

The review in this section has several limitations. Firstly, the mentioned theories only focus on social-psychological factors, while the environmental factors [52–54] on behavior are missing. Secondly, these theories are all based on the assumption that the behavior is intentional and conscious, while other researchers argue that supporting unconscious behavior change and habit fostering is more practical in preventative behavior interventions [7,55] because the most daily behavior is intuitive and automatic according to the dual-process theory [56]. Finally, we do not cover the implementation detail of each model and common pitfalls when using behavioral theories as discussed in [57].

Most of the interventions in the mentioned studies were delivered by face-to-face consulting, paper material, or email. The development of mobile technology, especially the mobile internet, has dramatically changed the way people get information and interact with each other. Therefore, it brings new chances to evaluate and improve the theories for the following reasons: 1) more in-situ data can be collected (e.g., measure the intention right before the potential behavior); 2) more ground-truth data can be gathered (e.g., GPS location and walking steps); 3) more unobtrusive intervention can be applied (e.g., providing the notification in proper time).

Following, we will discuss how to deploy the theory-based intervention elements with new technologies, especially mobile technologies.

# Technological Models

## Why Technological Models?

Technological models are introduced to guide the intervention implementation process using technology. An ideal technological model should provide a consistent taxonomy of intervention elements and a protocol to be followed when implementing interventions. The taxonomy enables formative reporting for systematic review. For example, we expect to design a mobile application to improve users' self-efficacy of physical activity. To this end, we may use task reduction (e.g., listing the possible running routes) and social learning (e.g., showing how the other people increase physical activity). The protocol can accelerate the implementation process and avoid neglecting necessary steps.

## Which Models?

We used the snowball method to obtain the highly related research starting from the work of Mohr et al. [58]. According to the relativeness and number of citations, we select the Behavior Change Techniques [31,59] and the Persuasive System Design and Behavior Change Support System [60,61] for the taxonomy, while the Behavioral Intervention Technology Model [58] and the Just-in-Time Adaptive Interventions [62,63] for the protocol.

## Taxonomies

### The Behavior Change Technique Taxonomy

BCTs are observable, replicable, and irreducible components of an intervention designed to change behavior [31,59], e.g., self-monitoring or goal setting. Abraham and Michie developed a taxonomy of behavior change techniques, which identified 22 BCTs and 4 BCT packages [31] and was later extended using a Delphi-type exercise, resulting in 93 BCTs clustered into 16 groups, called Behavior Change Technique Taxonomy (v1) [59]. The BCT taxonomy was intended to address some of the challenges posed by complex behavior change interventions, e.g., poor replicability due to a lack of consistency and consensus in reporting behavior change interventions and low comparability of the findings generated by systematic reviews due to the use of different systems for classifying behavioral interventions and synthesizing study results. It standardizes extracting active ingredients from reported intervention descriptions, helps to facilitate intervention development, design and reporting, as well as the investigation of possible mechanisms of action.

The BCT taxonomy has been used extensively, to inform intervention development and reporting [64,65], to systematically review the content of behavior change interventions and to identify BCTs associated with effectiveness [66–69], or to systematically appraise the potential of popular health and fitness apps [70–73] and wearables [74] that are currently available, to effectively assist health behavior change.

As increasing health interventions are delivered by mobile technologies, the mHealth holds great promise to enhance the evaluation of BCTs. For instance, based on the taxonomy, Belmon et al. [75] found that mobile phone Apps promoting physical activity, including the BCTs "goal setting and goal reviewing" or "feedback and self-monitoring", got higher ratings than those addressing "social support and social comparison" in a survey of Dutch young adults.

## The Persuasive System Design and the Behavior Change Support System

The concept of persuasive system design (PSD) and behavior change support system (BCSS) was proposed by Oinas-Kukkonen and Harjumaa [60,61]. Aimed at creating a conceptual framework that can be directly applied to persuasive system development, the PSD model describes the generic steps in persuasive system development and 28 principles in 4 categories (supporting primary task, computer-human dialogue, system credibility, and social). As discussed by Mohr et al. [58], the PSD model explains how to transfer these principles into software functionality. For example, to apply the reduction principle, the system should reduce users' effort to perform the target behavior. In the application of promoting healthy eating, a list of proper food choices can be provided. Besides, the Outcome/Change matrix in the BCSS model [61] is another strong contribution. It defines the type of change as attitude change (A-Change), behavior change (B-Change), and the change of an act of complying (C-Change), and the type of outcomes as forming (F-Outcome), altering (A-Outcome), and reinforcing (R-Outcome). Thus, more focused systematic review on PSDs is enabled.

We think the principles in the PSD model can serve as a taxonomy when designing mHealth interventions for two reasons: Firstly, it comprehensively describes how technologies can affect people. Secondly, there is evidence showing the effectiveness of the PSD principles. In 2012, Kelders et al. [76] provided a systematic review of the impact of the PSD on adherence to web-based interventions. Their model explained 55% of the variance in adherence in a hierarchical multiple linear regression and they found more extensive employment of dialogue support is related to better adherence. However, there were 14 other variables (e.g., study design) in addition to 3 category variables of PSD in their model, and the correlations of the PSD related variables and other variables were not reported. In [77], a meta-analysis shows that web-based interventions with the principles in the PSD model have a large and significant effect size on mental health, and increasing the number of principles in different categories does not necessarily lead to better outcomes. In addition, they also found several combinations of principles that were more effective, e.g., tunneling and tailoring, reminders and similarity, social learning and comparison.

## Protocols

### The Behavioral Intervention Technology Model

To provide a framework that can guide practitioners to design eHealth or mHealth interventions for behavior change goals, Mohr et al. [58] proposed the behavioral intervention technology (BIT) model in 2014. Using this model, researchers should be able to answer *why* they want to develop an intervention, *how* the behavior change will be achieved, using *what* technological element, and *how* and *when* it is technologically delivered. Given an exemplary treatment goal, physical activity promotion (the "why" part), one or more behavior change strategies should be chosen for the target group, for example, self-monitoring or motivation enhancement (the conceptual "how" part). During the instantiation section, BIT elements, e.g., passive data collection, visualization, and notifications, are designed (the "what" part) with different characteristics, e.g. personalization (the technical "how" part) and a specific workflow (the "when" part). Compared to the BCT taxonomy, the BIT model provides a logically complete set of components to be considered when designing the interventions. Although, as mentioned by the authors, the proposed model is a simplification as it is intended as a general framework. For instance, socialization and gamification are not included in the characteristics component of the BIT model.

Although the BIT model provides a practical protocol for designing behavior interventions, it is not without shortcomings. Firstly, the BIT model defines intervention aims as clinical aims and usage aims. In fact, we need to divide these aims into sub-aims or proximal outcomes, which is missing in the BIT model. Secondly, the conceptual "how" part of the BIT model refers to a list of behavioral intervention strategies, which are drawn only from BCT taxonomy [59], and we think including the PSD model can enrich this.

### Just-in-Time Adaptive Interventions Framework

In order to provide high-quality behavior change interventions, Spruijt-Metz argued that behavior change is a dynamic process varied over time, so interventions have to adapt to the users' needs in real-time [62,63]. The just-in-time adaptive interventions (JITAIs) framework describes a pragmatic protocol to guide intervention design, especially in the context of mHealth. The JITAIs framework requires researchers to define a distal outcome (e.g., increasing physical activity), proximal outcomes (e.g., daily walking goals), decision points (e.g., every two hours), intervention options (e.g., notification of the accumulated steps and "provide nothing"), tailoring variables (e.g., continuous sitting time), and decision rules (e.g., if continuous sitting time > 40 minutes, then provide a notification of the accumulated steps, otherwise, provide nothing). Tailoring variables are very important because they include the factors that determine vulnerability/opportunity and receptivity. For instance, of interventions to prevent binge drinking, psychological distress is the vulnerability/opportunity variable because it is the main predictor of drinking behavior, while whether the participant is driving or not is the receptivity variable because they cannot or should not receive any intervention when driving. Another property of the tailoring variables is timescale (e.g., the hourly process may have different variables from the daily process). The variables and decision rules give designers clear hints of what data to collect and how to implement the software workflow.

The JITAI emphasizes leveraging the context-aware ability of mobile technology to achieve the most important feature of mHealth intervention [62]. However, it is challenging to enable just-in-time interventions in some cases where context data is difficult to obtain. For example, a very common method to accurately assess individuals' sedentary behavior is to wear the sensor on the thigh, which may cause much inconvenience in daily life.

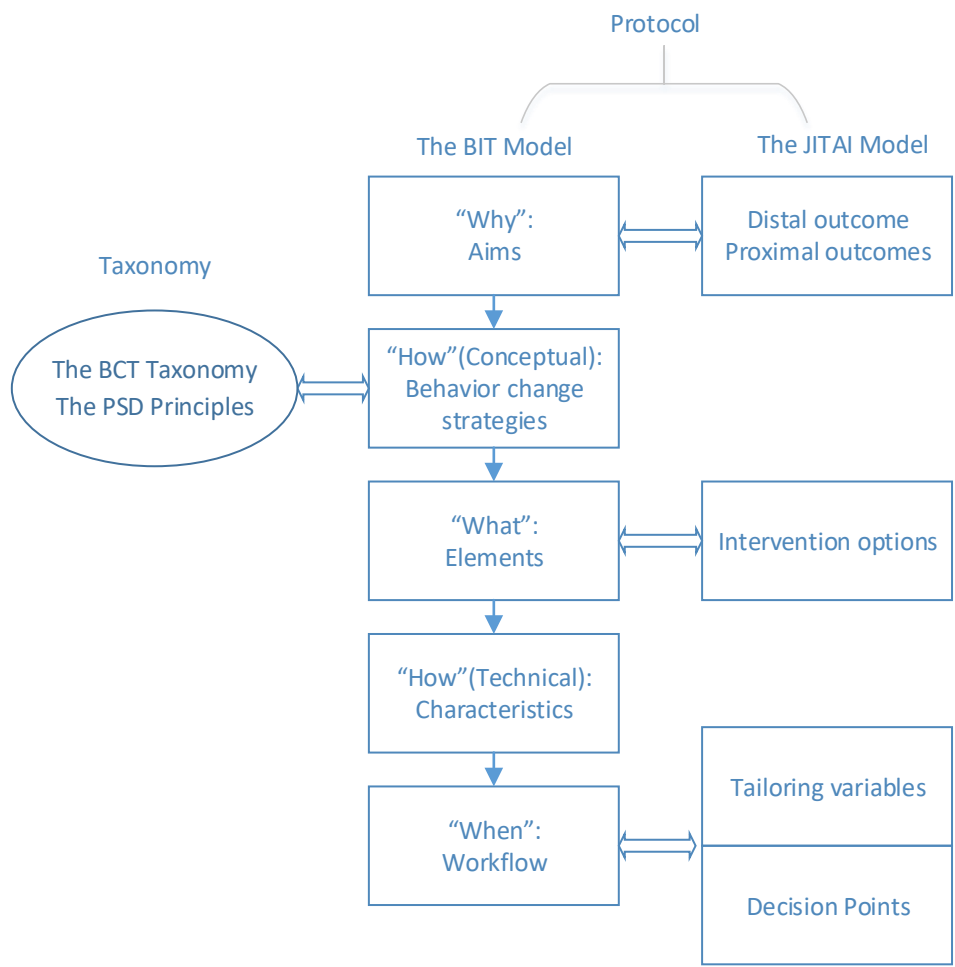

Figure 7: The technological model integrating the BCT taxonomy, the PSD principles, the BIT model, and the JITAI model.

## Discussion

We show a technological model integrating the BCT taxonomy, the PSD principles, the BIT model, and the JITAI model in Figure 7. We adopt the BIT model as the mainframe because of its generalization. The taxonomy of BCTs and the principles of PSD can be served as repositories of intervention strategies. These definitions in the taxonomy should be followed by related studies so that more evidence can be accumulated for more powerful conclusions. In addition, we suggest explicitly reporting the target factors based on behavioral theories (e.g., perceived benefits) to reveal the correlation between them [78]. The advantages of JITAIs can enrich the BIT model, especially in the workflow part.

## Systematic Approaches

We have discussed the behavioral theories and technological models. Now we will introduce systematic approaches based on user-centered design process to link the two parts together. Unlike laboratory-based experimental studies, mHealth intervention studies always need to be conducted in the wild, long-term. A small mistake may cause catastrophic results. For example, a mobile application is designed to compare the effect of two different visualization methods on motivating users to achieve their daily physical activity goal. Users may drop the app simply because it is battery draining or it requires manual logging to collect data. To maximize the chance of a successful study, user-centered design ought to be followed.

In this section, we analyze three systematic approaches, the CeHRes Roadmap [6], Wendel's approach [7], and the IDEAS model [79], illustrating the important factors commonly emphasized. Though different terms are used in the mentioned models [6,7,79], the main phases or workflows are similar according to their descriptions, as shown in Figure 8. For the sake of consistency, we define the phases as *understanding, targeting, designing, evaluating, and refining.*

*Understanding the human mind, behavior change, and the problem*. The behavioral theories we reviewed in this paper provided us with the tools to understand why and how people change. However, the understanding should not be limited to the mentioned theories. For example, the dual-process theories [80] tells that there are two distinct systems in human mind: conscious and unconscious. Though there are debates on the definition and features of dual-process theories [81], we can still learn the basic idea from it, that many of our daily behaviors are done unconsciously. Given this knowledge, we can tell why people sometimes make wrong decisions without being aware of it and improve the intervention by adding triggers and cues accordingly. In addition to the human cognition, the understanding of the target problem is equally important. For instance, the prevalence of myopia in Chinese children is one of the most concerned health problems in China. Some parents think the near-work activities, such as reading books, are the main reason of myopia. However, a study showed that the prevalence of myopia in 6- and 7-year-old children of Chinese ethnicity was significantly lower in Sydney than in Singapore and the time on outdoor activities was the most significant factor, and not near work activity [82]. Given the fact that chronic diseases are strongly correlated with bad habits or unhealthy routines [83], understanding the problem in a holistic manner is necessary [11].

*Targeting the user group, the behavior, and the outcome*. Only when the target group, the target behavior, and the expected outcome are clear, can we design the most promising interventions. To reduce sedentary behavior, for example, interventions should be designed separately for young people and old people, because of their different ability. The outcomes determine the measurement, which is related to user burden. The tradeoff between the measurement and user burden should be noticed when deciding outcomes.

*Designing the intervention*. The technological models discussed in the previous section should be applied in the design phase. If inter-disciplinary groups are involved, they should thoroughly communicate about the tasks for each group as well as implementation details. A clear requirement document can greatly increase cooperation efficiency.

*Evaluating the product or the solution*. Researchers from different background hold different opinions about evaluation. Randomized controlled trials (RCTs) are the golden standard in many scientific fields, especially medical and healthcare. However, methods beyond the RCT are also suggested when applying the mHealth intervention to complex environment [6]. One key factor for evaluating mHealth interventions is to make the measurement as direct as possible, by which clear evaluation results can be drawn. On the other hand, [5] suggested adopting factorial ANOVA experimental design for evaluating the power of intervention components based on the advantage of e-health interventions, which offers low cost to deliver multi-condition interventions.

*Refining the design*. According to the feedback of the users, refinement may be conducted to improve the product. This step can be iterated many times with prototyping and piloting. The refinement should be based on a proper understanding of the feedback. For example, it makes sense to get feedback only in real life scenario when designing a JITAI for binge drinking interventions.

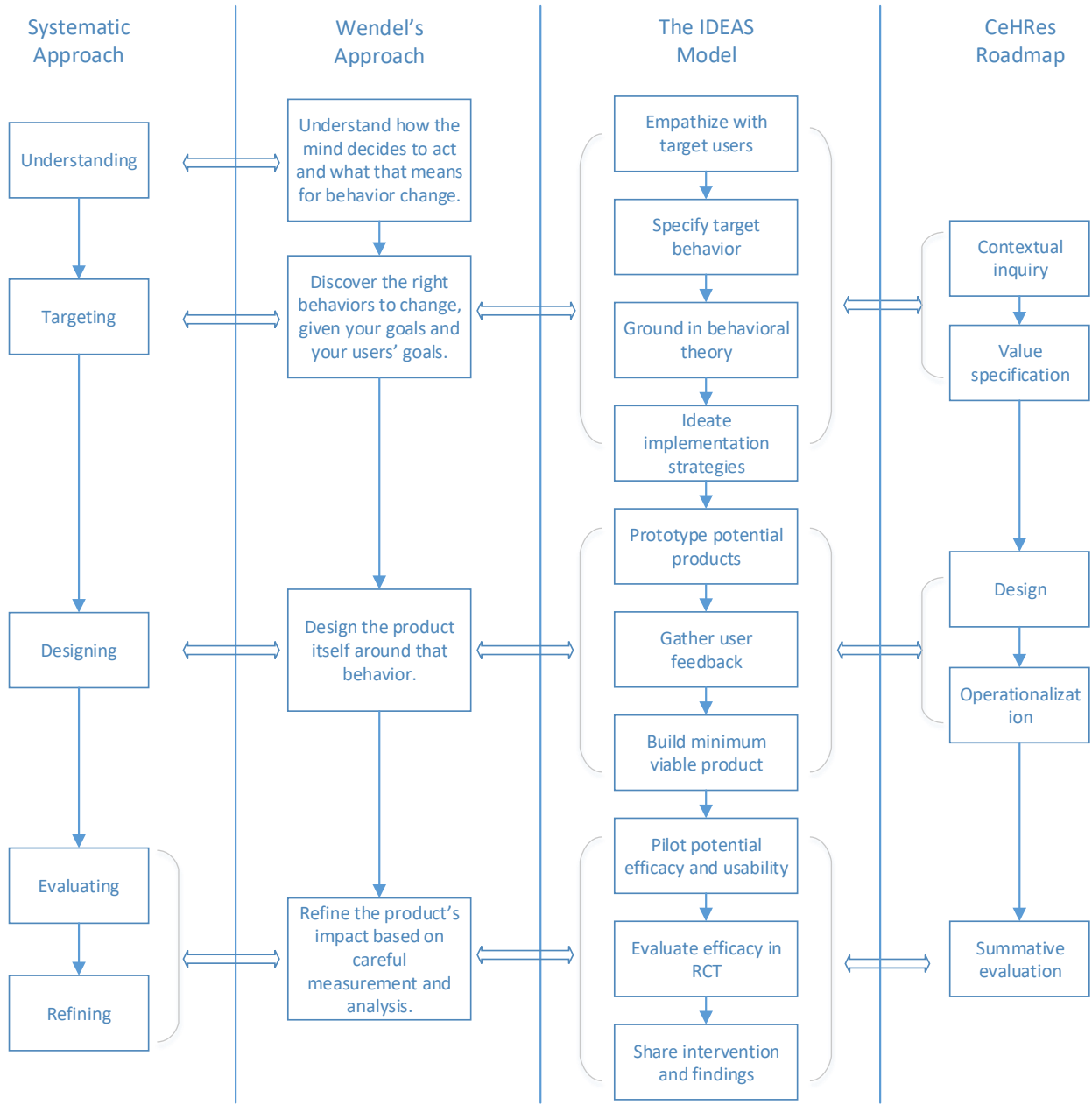

Figure 8: The comparison of Wendel's approach, the IDEAS model, and the CeHRes Roadmap, adapted from [7], [79], and [6], respectively.

We compare the workflows of Wendel's approach, the IDEAS model, and the CeHRes Roadmap in Figure 8. Most of the steps are common among these approaches. Wendel's approach covers all the key phases we discussed, while the only logic difference is that we divide the last phase in Wendel's approach to evaluating and refining. We separate these two phases because they are not always coupled, i.e., in some experiment designs the workflow may stop at the evaluation phase because the evaluation result meets the expected outcome.

## Example

After the illustration of three categories of theories and models within our holistic approach, we will demonstrate how to apply this approach by discussing a published mHealth intervention research which aims to reduce office workers' sedentary behavior [84]. This research included two experiments covering most elements in our holistic approach, while it lacked some imperative steps that could have been improved.

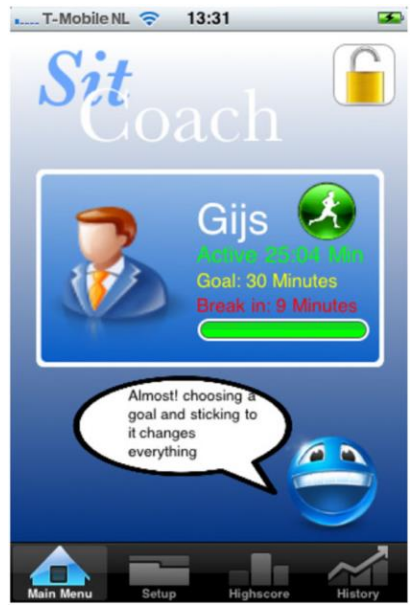

Figure 9: The SitCoach from [84].

First, the authors introduced the detrimental effect of prolonged sitting and benefit of taking breaks from sitting. This is the *understanding* of the problem. The sedentary awareness (i.e., perceived severity) and the perceived behavioral control are the constructs from behavioral theories that were considered. The SitCoach (shown in Figure 9), a mobile application, was developed in experiment 1 *targeting* increasing sedentary awareness and reducing sedentary behavior for office workers with high computer dependence. The outcomes of this experiment are the usability of the mobile application, and users perceived behavioral control in changing sedentary behavior. The behavior change techniques adopted in the phase of *designing* included self-monitoring and reminders. The reminder is timely (i.e., after a certain period of sitting) and multi-modal (i.e., vibration and buzzing). After the one-day experiment, an *evaluation* of interviewing the participants with two questionnaires was conducted and it was found that participants were not aware of the harmfulness of sedentary behavior and they had little perceived control over their sitting behavior. Furthermore, the demand for carrying and recharging the dedicated smartphone with the application annoyed the participants. Based on the lessons learned from the first experiment, a follow up experiment was conducted.

*Table 1: the corresponding elements in the experiments from [84].*

| | | Experiment 1 | Experiment 2 |
|---|---|---|---|
| Understanding | Problem | Prolonged sedentary behavior relates metabolic syndrome; the risk of sedentary behavior is independent of one's overall physical activity level | Same as experiment 1 |
| | Related constructs from behavioral theories | Perceived severity; perceived behavioral control | Subjective norm; perceived severity; perceived behavioral control |
| Targeting | Behavior | Sedentary behavior | Sedentary behavior |
| | People group | Office workers with high computer dependability | Office works with a predominantly sedentary job |
| | Outcomes | Usability of the intervention tool; perceived control of the behavior | Computer activity; physical activity |
| Designing | Behavior change techniques | Self-monitoring; reminder | Self-monitoring; reminder |
| | Persuasive strategies | No | Authority; commitment; liking; reciprocity; scarcity |
| | Interaction methods | Data collection: smartphone reminder: vibration, buzzing, and visual information on a smartphone | Data collection: using dedicated activity tracker to record physical activity; using computer software to record computer activity reminder: SMS |
| Evaluating | Study setting | Feasibility study (8 people, 1 working day) | Controlled experiment (86 people, 7 weeks) |
| | Tools and measurements | Using questionnaires to evaluate the usability of the intervention tool and the perceived behavioral control of the users | Passively collecting physical activity, computer activity, and user's reaction to the SMS |

|  | Results | Results show that the participants have little perceived severity of sedentary behavior and have low perceived control over their sitting behavior; Vibration is the most preferred reminder modal and battery train is disturbing | Results show that timely messages can effectively reduce computer activity, physical activity appeared to peak immediately after a persuasive message in the intervention group, and no additional effect by persuasive strategies |
| --- | --- | --- | --- |
| Refining |  | See the second experiment | Further intervention design could increase sample size, duration, and other outcomes |

In the second experiment, some *refinement* was made in each phase. In the *understanding* phase, subjective norm was considered, while in the targeting phase, subjective measurements (i.e., physical activity and computer activity) were added. Furthermore, persuasive strategies (i.e., authority, commitment, liking, consensus, and scarcity) were used to create the reminder (i.e., a short text message with a hyperlink) to increase the intervention effectiveness. The results showed significant effectiveness in reducing computer activity and increasing calories burned. More details about the corresponding elements are shown in Table 1.

These two experiments were well designed in a user-centered manner and can be clearly mapped into the process of *understanding, targeting, designing, and evaluation*. By using our holistic approach, we can find the participants were not assessed using behavioral theories, e.g., comparing the perceived severity and perceived behavioral control over their sedentary behavior in baseline and post-intervention assessments. Despite the limitations of this example, it shows how to use our proposed holistic approach step by step in real mHealth intervention applications.

## Limitations

This paper has several limitations. Firstly, as our holistic approach aims to provide a structural guidance of the process of designing mHealth interventions, it is a simplification. Secondly, we select theories and models based on related review papers and the authors' own literature retrieval, but not from a systematic review. Thirdly, when comparing models, we abstract the aspects that are necessary for our analysis, but not cover all the facets of each model. Lastly, the example we use to illustrate the process of applying the proposed approach is derived from a related work, not our own experience. We encourage the elaboration and refinement of this holistic approach.

## Conclusion

In this paper, we provided an additional tool for designing mHealth interventions. Based on our holistic approach, behavioral theories should be embedded into the user-centered design process when developing mobile health interventions. Evidence shows that there is no "one-size-fits-all" behavioral theory. Therefore, it is suggested to utilize several theories to cover all potential constructs affecting a specific behavior. Based on the Behavior Intervention Technology Model and the Just-In-Time Adaptive

Interventions framework, we developed a protocol combined with the Behavior Change taxonomy and the Persuasive System Design principles. Finally, we abstracted the common steps of the user-centered design process from three existing approaches. Although the proposed approach targets mobile health interventions development and the example given is unique to mHealth intervention development, the approach can generally be applied to other health intervention with some restrictions to real-time adaptive intervention techniques.

There are still several open questions when using this holistic approach. We have alluded to several, including: (1) How to match constructs from behavioral theories to behavior change techniques? (2) How to model and detect human behavior to be aware of the context of interventions? (3) How to define the measurement used to evaluate mHealth interventions? These questions will be explored as part of our future work.